\documentclass[a4paper,11pt]{article}

\usepackage{jinstpub} 
\usepackage{subfig}

\title{\boldmath A Method To Characterize Metalenses For Light Collection Applications}

\author[a,1]{T. Contreras,\note{Corresponding author.}}
\author[b]{A. Martins}
\author[c]{C. Stanford}
\author[c]{C. O. Escobar}
\author[b]{R. Guenette}
\author[c]{M. Stancari}
\author[d]{J. Mart\'{i}n-Albo}
\author[c]{B. Lawrence-Sanderson}
\author[c]{A. Para}
\author[c]{A. Kish}
\author[d]{F. Kellerer}

\affiliation[a]{Department of Physics, Harvard University, Cambridge, MA 02138, U.S.A.}
\affiliation[c]{Fermi National Accelerator Laboratory, Batavia, IL 60510, U.S.A}
\affiliation[b]{Department of Physics, University of Manchester, Manchester M13 9PL, United Kingdom}
\affiliation[d]{Instituto de F\'isica Corpuscular (IFIC), CSIC \& Universitat de Val\`encia, Calle Catedr\'atico Jos\'e Beltr\'an, 2, Paterna, E-46980, Spain}

\emailAdd{taylorcontreras@g.harvard.edu}

\abstract{Metalenses and metasurfaces are promising emerging technologies that could improve light collection in light collection detectors, concentrating light on small area photodetectors such as silicon photomultipliers. Here we present a detailed method to characterize metalenses to assess their efficiency at concentrating monochromatic light coming from a wide range of incidence angles, not taking into account their imaging quality.}


\begin{document}
\maketitle
\flushbottom

\section{Motivation}
\label{sec:intro}

Light collection is essential for detectors in a wide range of applications. Future particle physics detectors looking for dark matter, studying neutrino interactions and oscillations, or searching for neutrinoless double beta decay, would significantly benefit from increased  light collection to reach, and even enhance, their physics goals. Other applications for light collection detectors include medical physics like positron emission tomography, material science using photochemistry and mass spectroscopy, and photonics using non-linear optics. 

\begin{figure}[htbp]
\centering 
\subfloat[]{\includegraphics[width=.5\textwidth]{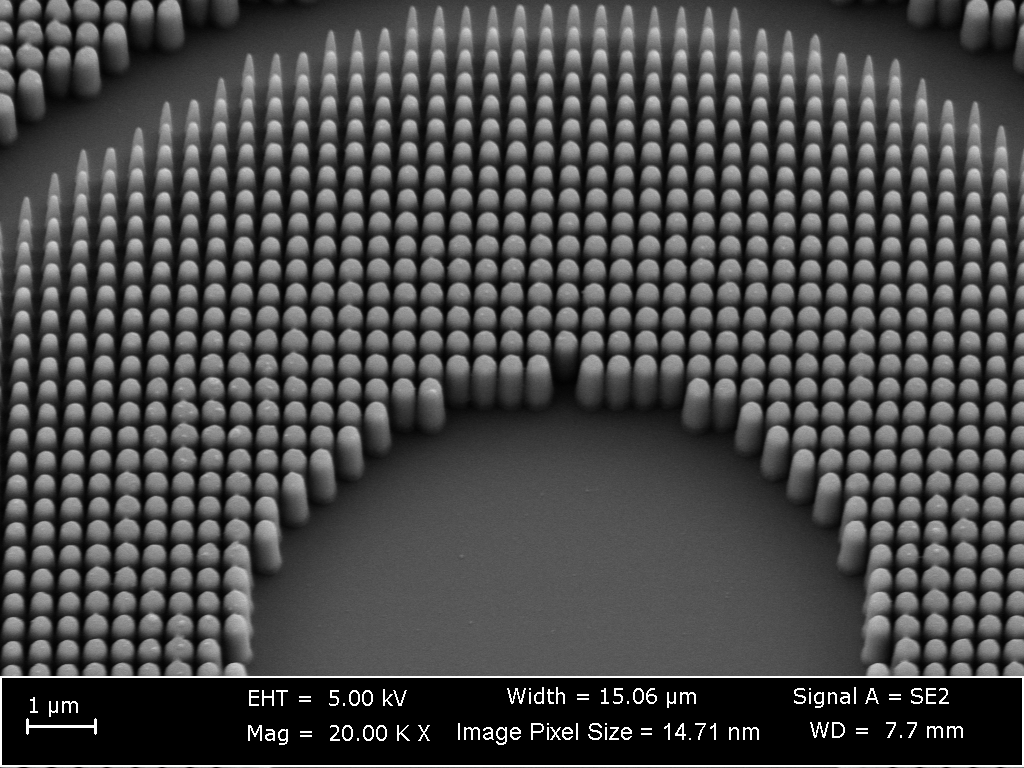}\label{fig:sem}}
\subfloat[]{\includegraphics[width=.375\textwidth,trim={5cm 20cm 0 20cm},clip]{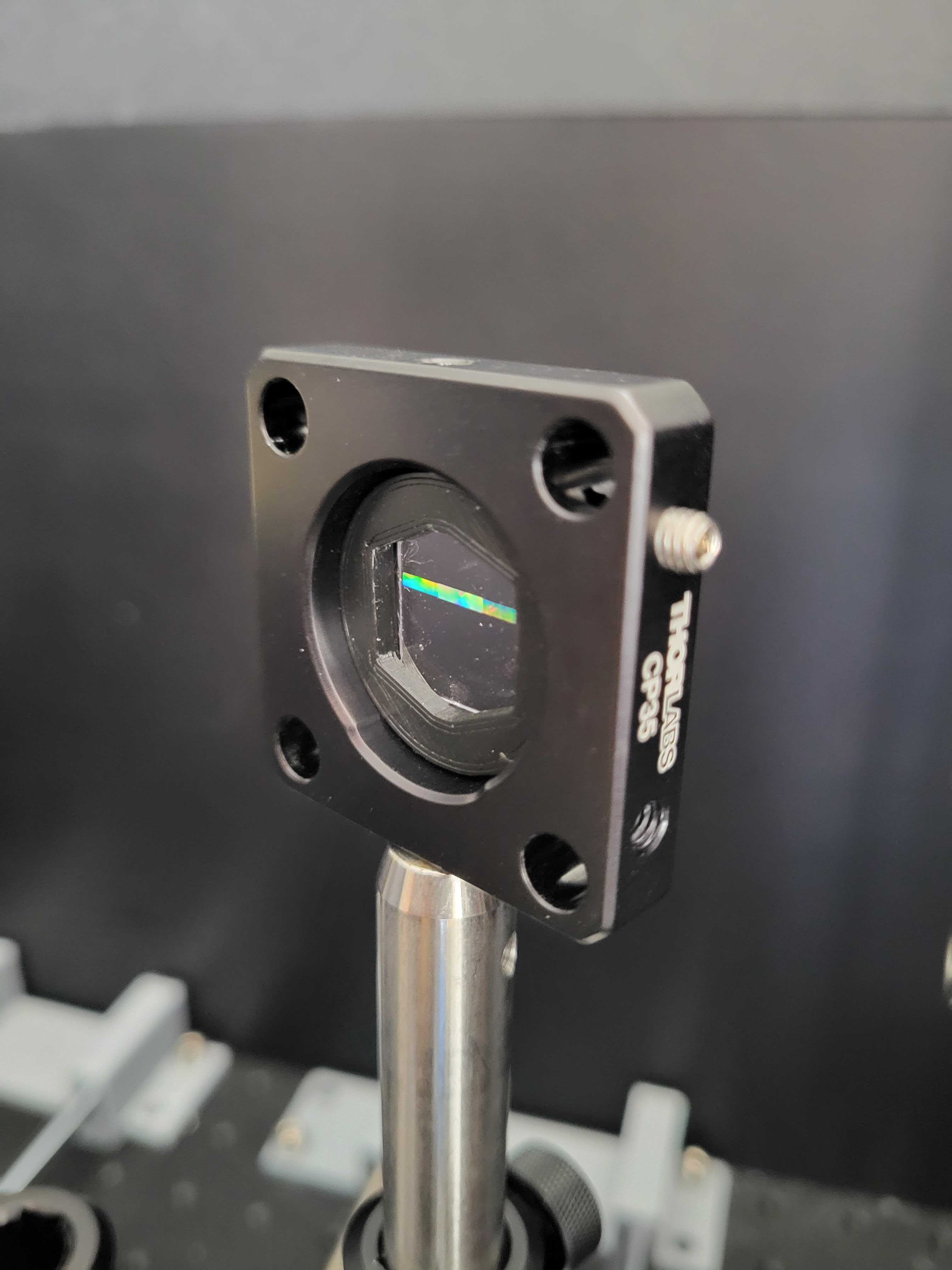}\label{fig:lens}}
\caption{ (a) A scanning electron microscope micrograph of a metalenses designed for 632 nm used for the characterization. The arrangement of nanopillars of different sizes (from 150 nm to 460 nm) can be seen.  (b) A metalens mounted to be characterized. This metalens was fabricated with only a thin stripe across, for simplicity in characterization and fabrication, with the understanding that the metalens would be circularly symmetric. }
\end{figure}

Metalenses, composed of dielectric nanostructures that focus light like a lens, shown in Figure \ref{fig:sem}, are a novel and versatile technology that could greatly increase the efficiency of light collection in detectors while being small, practical, durable, radiopure, and cost-effective \cite{Mohammadreza2017,Liang2019}. The nanostructures in a metalens can be made out of a variety of materials and shapes of subwavelength dimensions, allowing for a wide range of specific optical properties. These devices are compatible with mass manufacturable processes that are used in the chip industry, which enables their fabrication in large scales, making metalenses cost-effective compared to traditional lenses \cite{Park2022,Zhang2023,Kim2023,Park2019}. 

With this versatility, metalenses could be paired with light detectors in many ways to improve light collection. In previous work, we demonstrated that metalenses can be an attractive solution to increase the light collection of small photon detectors by a factor of 7, despite using a non-optimized metalens design \cite{LoyaVillalpando2020}. The use of metalenses in large-scale detectors could provide a solution to increasing the light collection without increasing the number of photon detecting channels. Metalenses could also be paired with silicon photomulitpliers (SiPMs) or CMOS devices to improve the photon detection efficiency \cite{Uenoyama2021,Mikheeva2020,Enoch_2021}. 

Metalens development has mainly focused on imaging purposes \cite{Mohammadreza2017,Liang2019,Martins2020,Mohammadreza2016,Pahlevaninezhad2018,Park2019}, where the attention was put on the quality of the images produced from fully illuminating the lenses rather than on the efficiencies of transmitting and focusing the light for different parameters. When investigating a metalens for high-efficiency light focusing, the characterization of the optical properties need to be done for parameters such as incident angle, distance from the lens, outgoing angle, and field of view. This is also essential to allow detailed simulations of the impact of metalenses in a large-scale detector. 

In this paper, we present a method to characterize metalenses for light collection, rather than image quality, for the development of metalenses for light collection detectors. By measuring the efficiency for a range of incident angles of light, position from the lens, and angles of outgoing light, simulations of specific detector geometries and light profiles can accurately estimate light collection improvements when incorporating metalenses into light detector designs. We begin this development with an optimized metalens designed for 632 nm light and compare it to the simulated response. This wavelength is easier to fabricate and to test this method of characterization, which can then be used with minimal changes for metalenses of other wavelengths in future studies.

\section{Description of the metalens used for the characterization}

The metalenses are made of fused silica nanoposts combining two design approaches, one for the low deflection angle region (phase map design approach) and one for the higher deflection angle region (supercell design approach), to optimize the efficiency in each region. The fabrication was performed using lift-off based electron-beam lithography \cite{Park2019}. 

\begin{figure}[htbp]
\centering 
\includegraphics[width=.7\textwidth]{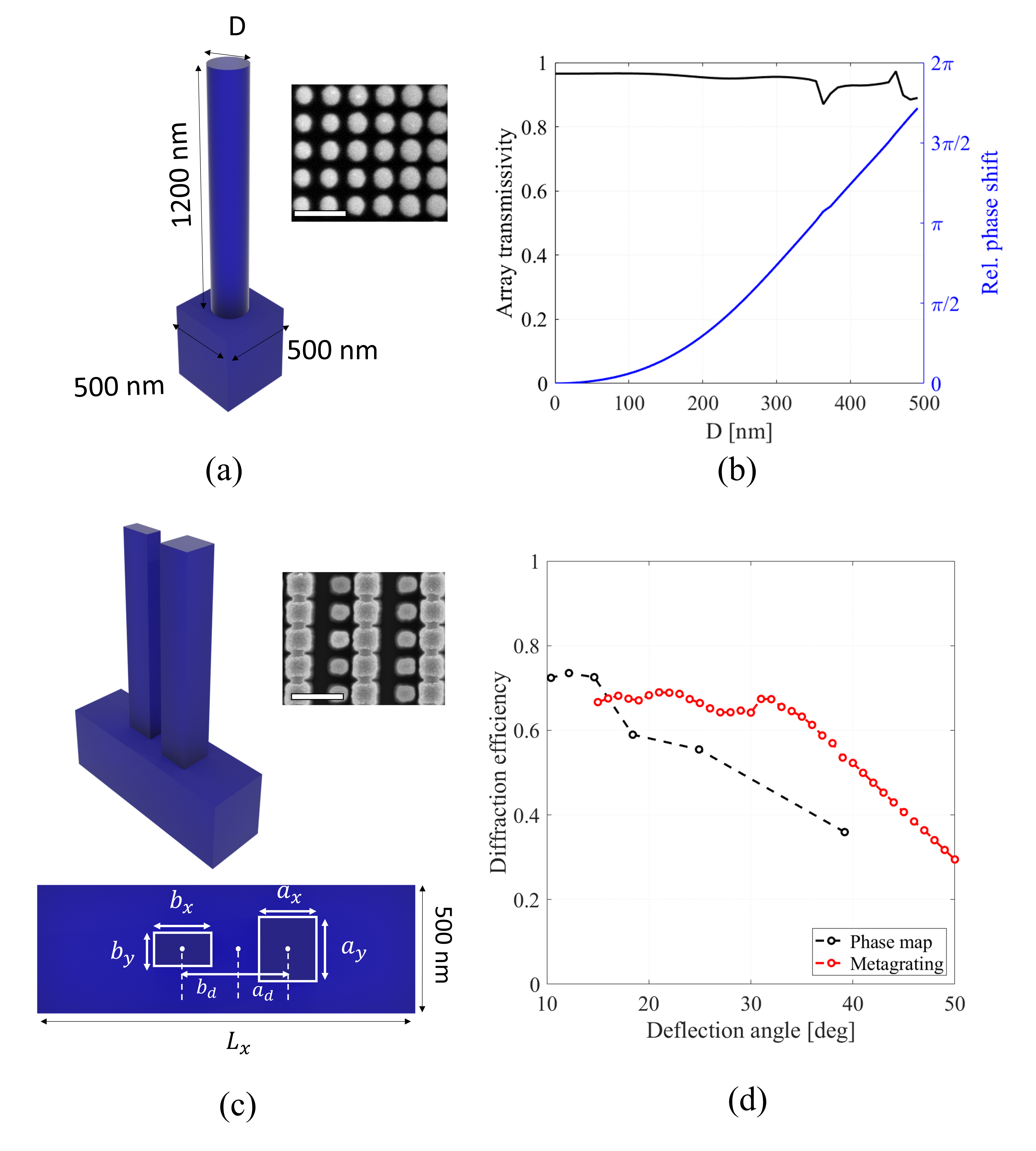}
\centering
\caption{\label{fig:meta}  (a) Phase map design based on 1200 nm tall silica nanoposts. The unit cell is fixed at 500 nm and the structure is designed to operate at 632 nm. The insets in (a) and (c) show SEM micrographs of some of the corresponding fabricated structures. The scale bar represents 1 $\mu$m. (b) Transmission and relative phase shift of the nanoposts  in (a) as function of their diameter. In the design, we constrained the minimum diameter to 150 nm. (c) Optimized silica-based metagrating structure, which uses a pair of coupled nanoposts. The inset shows a top view  of the structure with the optimized parameters. The long period $L_x$ was fixed according to the diffraction equation for each deflection angle, and the small period was fixed at 500 nm in all cases. The operating wavelength is 632 nm. (d) Diffraction efficiency of the supercell design metagratings (red dots) and of beam steering metasurfaces designed with the phase map approach (black dots) shown in (a) and (b). }
\end{figure}

In the low deflection angle region (less than $\approx$20$^{\circ}$), we used the conventional phase map design approach to modulate the transmitted field with a hyperbolic phase profile \cite{Park2019}. We designed 1200 nm tall nanoposts on silica with a unit cell size of 500 nm using the phase modulation design, as shown in Figure \ref{fig:meta} (a). Figure \ref{fig:meta} (b) shows the post array phase shift and transmittance as function of the post diameter used to encode the metalens phase profile. However, the phase map design has a small efficiency when it is used to steer light at higher deflection angles due to the excitation of higher order modes when the array symmetry is broken \cite{lalanne2017metalenses}. In other words, for high deflection angles (greater than $\approx$20$^{\circ}$) the local approximation assumed in this design is no longer valid and the array phase shift of each post does not correspond to the phase shift it can provide in the metasurface or metalens. This is evident in the black dots shown in Figure \ref{fig:meta} (d) that shows the deflection efficiency of beam steering metasurfaces designed using the phase control method of Figures \ref{fig:meta} (a) and (b). Since the conventional nanopost period is fixed at 500 nm, only supercells with unit cells that are multiples of 500 nm were calculated. Note that the efficiency drops for deflection angles higher than 20$^{\circ}$, approximately.   

Such reduction of efficiency at high deflection angles would limit the numerical aperture (NA) and overall efficiency of our metalens. Thus, to avoid this issue, we used the supercell design approach with metagratings to steer light at high deflection angle regions \cite{Paniagua2018}.  The metagratings are optimized to maximize the deflection efficiency at a certain angle using more than one post per supercell or with a freeform shape \cite{fan2020freeform}. Our design consists of two 1200 nm tall glass coupled nanoposts with rectangular cross-section, as shown in Figure \ref{fig:meta} (c). We fixed the orthogonal period at 500 nm, which is subwavelength in air for a 632 nm excitation wavelength. The longer period, $L_x$ was chosen using the grating equation to diffract at an angle $\theta$ as shown in equation \ref{eqn:gratingeq}
\begin{equation}
\label{eqn:gratingeq}
   L_x = \frac{\lambda_0}{\sin{\theta}}
\end{equation}

We used a genetic algorithm to optimize the first order efficiency by varying the remaining geometrical parameters, shown in Figure \ref{fig:meta} (c). We constrained the minimum gap between the posts at a minimum value of 50 nm and the minimum post size at 200 nm. All calculations were performed using the rigorous coupled wave analysis method (RCWA) \cite{Popov2001,Whittaker}. The red dots in Figure \ref{fig:meta} (d) show the resulting first order efficiency of the optimized structures using the supercell design approach with metagratings as a function of the deflection angle. The efficiency of the meta gratings remain higher than ~65\% for diffraction angles as high as ~32$^{\circ}$, which corresponds to  $\sin \theta \approx$  ~0.5. On the contrary, the phase control design efficiency remains higher than 70\% only for deflection angles smaller than ~20$^{\circ}$ and quickly reduces for higher deflection angles. It should be noted that, the design of metagratings is not practical for small deflection angles as their periods are inversely proportional to the sine of the deflection angle, according to equation \ref{eqn:gratingeq}. Therefore, in our metalens design we used the phase control design for deflection angles $\theta$ up to 20$^{\circ}$ of deflection, corresponding to  $\sin \theta$ < 0.36, and the supercell design for deflection angles larger than 20$^{\circ}$, corresponding to  0.36 < $\sin \theta$ < 0.75, resulting in a metalens optimized for efficient first order deflection across a wide range of deflection angles. 

The data presented here were taken with a rectangular metalens of 1.2 mm width and 13.7 mm length, with a focal length of 7 mm and a NA of 0.75, combining the two designs described and placed on a glass substrate. The optimized metalens was produced as a rectangle (shown in Figure \ref{fig:lens} rather than a circular metalens to increase speed of prototype production and testing, and considering a symmetric response can be easily obtained in electron-beam and photolithographic processes. The width of the rectangular metalens was determined to completely enclose the width of the light beam, described in section 3. We simulated a 300 $\mu$m Gaussian beam reaching a metalens on different positions. Our model can accurately simulate a large area metalens in a timely manner.

\section{Characterization methodology and experimental setup}

When studying the application of metalenses to particle physics detectors, which have a wide range of geometries and photon detection arrangements, it is important to have a detailed characterization of the optical components as a function of different incoming angles and distances between the lens and the photodetectors. This will provide the necessary optical parameters needed for detailed particle physics detector simulations. 

\begin{figure}[htbp]
\centering 
\includegraphics[width=.6\textwidth,clip]{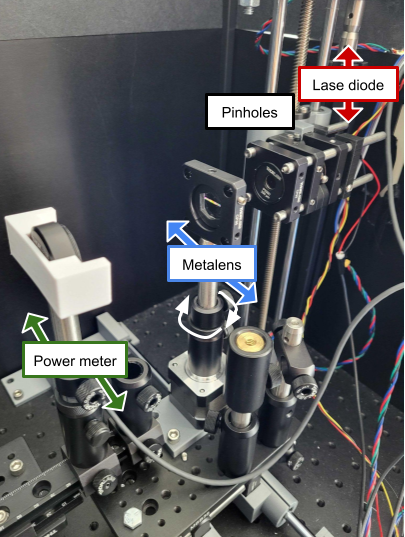}
\caption{\label{fig:darkbox} Experimental setup developed for the metalens characterization with the laser diode on the far right, followed by the two pinholes, with the metalens in the center and a power meter on the left. Arrows next to labels of the devices in the experiment demonstrate the direction the motors can move the given device.}
\end{figure}

\begin{figure}[htbp]
\centering 
\includegraphics[width=.9\textwidth]{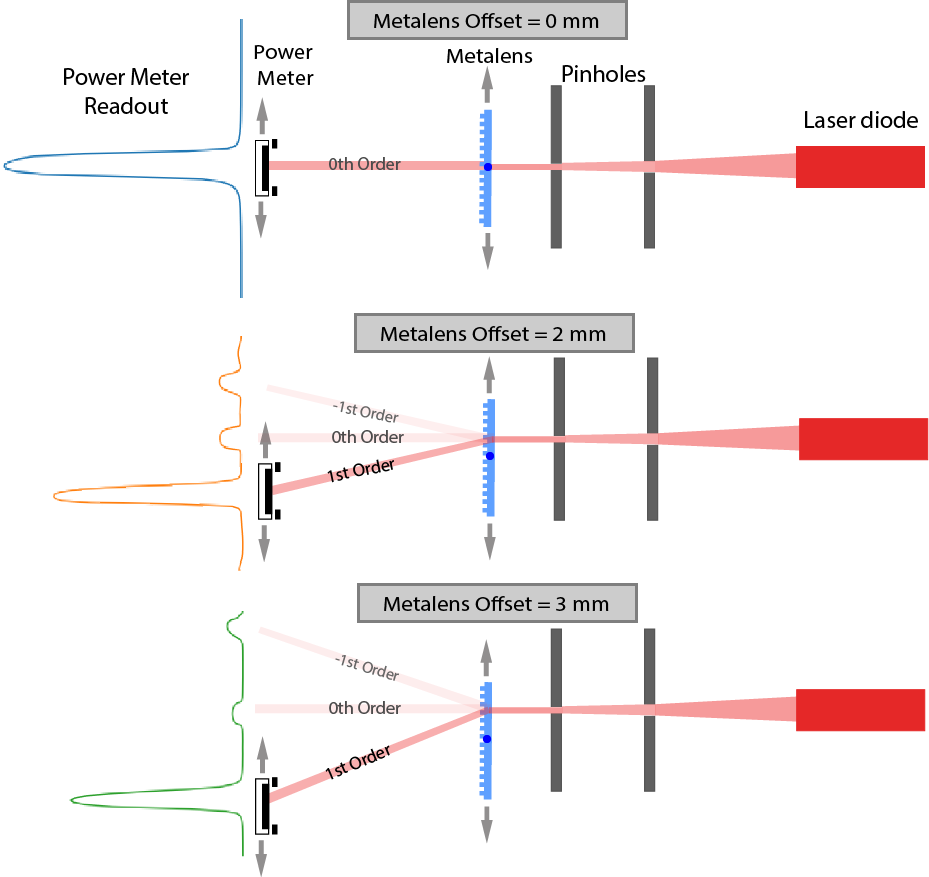}
\centering
\caption{\label{fig:diagram} Diagram of the experimental setup and power meter output for three different positions of the light source relative to the metalens. The light source (laser diode) is collimated by two pinholes forming a beam spot size of ~300 $\mu$m on the metalens, which focuses the light onto a photon detection device (power meter). An adjustable slit is in front of the photon detection device, which is used for alignments and left completely open for data taking. The light path is illustrated in red, where the 0th and 1st diffraction order of the metalens are shown.}
\end{figure}

In order to fully characterize a metalens, we built an experimental setup capable of automatically moving a sensor, a metalens, and a light source linearly in addition to rotating the metalens to change the angle of incidence of the light on the metalens. The setup can be seen in Figure \ref{fig:darkbox} and is illustrated in Figure \ref{fig:diagram}.

A 3 mm $\times$ 3 mm S13370 \textsc{Hamamatsu} VUV SiPM paired with a red 632 nm LED was used initially, as SiPMs can detect single photons and are often used in noble element detectors. The VUV SiPM has sensitivity up to 900 nm, thus could still be paired with the red LED.\footnote{\url{https://hamamatsu.su/files/uploads/pdf/3_mppc/s13370_vuv4-mppc_b_(1).pdf}} To compare the results, we also used a \textsc{Thorlabs} S120SV Power Meter paired with the 632 nm laser diode to provide sufficient power detectable by the power meter. The power meter and laser diode setup allowed us to get a smaller spot size on the metalens. We found the results comparable between the SiPM and power meter setup, and thus continued using the power meter setup as we optimized the metalens design due to the increases speed of data taking with the power meter. As such, data presented here was taken only with the power meter setup. 

The light source, at one end of a dark box, is coupled to two pinholes, 200 $\mu$m and 400 $\mu$m, to create a beam of light with a spot size of approximately 300 $\mu$m on the metalens. The metalens then redirects, or focuses, light to the plane with the sensor paired with an adjustable slit shown in Figure \ref{fig:diagram}. The slit allows for alignment of the system between the sensor center and the beam.

The experimental setup includes \textsc{PRUSA} stepper motors connected to \textsc{Thorlabs} optics boards with 3D printed pieces for mounting.\footnote{\url{https://www.prusa3d.com/}} Three motors are assigned to linearly move the sensor, metalens, and light source, and one motor to rotate the metalens. Each motor is connected to its own easy driver, with all motors connected to one \textsc{Arduino} board. The stepper motors allow for fine-grained linear movement down to 0.1 mm and angle rotations down to 0.225 degrees. A Python-based software package was developed to easily control each motor, by using the \textsc{PyFirmata} package to connect to the \textsc{Arduino} board. The developed software package includes automatic alignments of the system of sensor, metalens, and light source as well as the software for data taking. With this software, we are able to move all motors and measure the power from the power meter remotely. This allowed us to completely align the system automatically, by scanning across the lens to find the center of the metalens as well as the angle relative to the light beam. While aligning, continuously updating plots give a live view of the varying power of the power meter as it is moved perpendicular to the light beam. 

The slit is only used for alignment, as the slit cannot be placed close enough to the power meter such that it does not block the light reaching the power meter at higher deflection angles. The slit is left open after aligning and while taking data with the power meter. \textsc{Thorlabs} Optical Power Monitor monitors the power meter, and was used to measure the average power from the metalens-laser diode beam by averaging 15 power measurements found over 0.15 seconds.

\begin{figure}[htbp]
\centering 
\subfloat[]{\includegraphics[width=.55\textwidth]{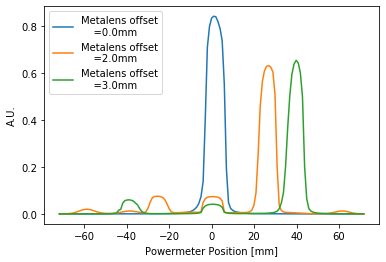}}
\subfloat[]{\includegraphics[width=.4\textwidth,origin=c]{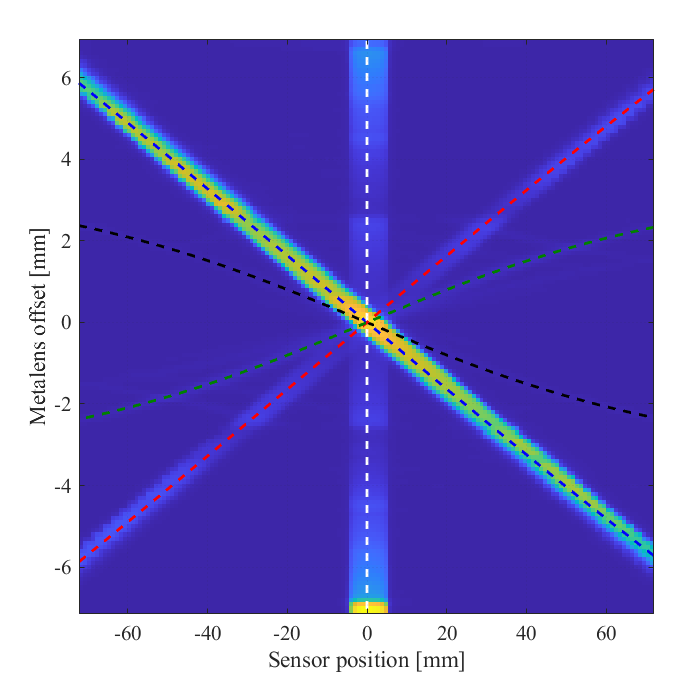}}
\centering
\caption{\label{fig:waveforms} (a) Waveforms showing the light intensity seen by the power meter, with zero representing the power meter centered with the light source. Each waveform shows the response of the metalens when offset relative to the light beam spot (metalens and light beam centered in blue, metalens offset 2 mm from center in orange, 3 mm from center in green). The zeroth order, where the light goes straight through the lens, is at the 0 position, while the first order shifts based on the position of the light source on the metalens. (b) A light map of the entire data set. The first order (m= -1 , focusing) is clearly seen as the brightest band at an oblique angle, while the zeroth order is the vertical line. The m=+1 order (diverging) can be seen as the lighter tilted band. The overlay lines show the expected positions for the diffraction orders according to the grating equations \ref{eqn:gratingeq}.}
\end{figure}

While taking data, the light source stayed in place, focused by the pinholes to approximately 300 $\mu$m on the metalens, creating a diffraction pattern on a perpendicular plane to the lens. The power meter and metalens then move in small steps of 0.2 mm and 1mm, respectively, across this light pattern. The zeroth order of this pattern is the light going straight through the metalens. The first order is the desired angle in which the metalens is optimized, as shown in Figure \ref{fig:waveforms} where the first order is the largest peak of each waveform, and the deflection angle is determined from the power meter position combined with the metalens offset relative to the light beam. A reference measurement, used to normalize the metalens output, was taken with the light going straight through the bare glass substrate. 

The orders were determined by their distance from the zeroth order, which is always centered at zero, shown in Figure \ref{fig:waveforms}. The peak position of the m-th diffraction order on the SiPM plane ($d_{m}$)  can be found by 
\begin{equation}
    \label{eqn:difraction_orders}
    d_{m} = d_0\tan{(\theta(r,m)+\theta_0)}
\end{equation}

where $d_0=10 cm$ is the distance between the power meter and the metalens,   $\theta_0$ is the rotation angle of the metalens, r is the metalens offset and $\theta(r,m)$ is the diffraction angle, which for a focal length $f$ is given by \cite{doi:Capasso1}

\begin{equation}
    \label{eqn:diffraction angle}
    \theta(r,m) = \arcsin{(\frac{mr}{\sqrt{r^2+f^2}}-\sin\theta_0)}  
\end{equation}

\section{Results}

The response of the metalens for various diffraction orders can be seen in Figure \ref{fig:waveforms}a, with each line representing a different metalens offsets relative to the light beam. With the light beam centered on the metalens, we get only the zeroth order, shown in blue. As we move the metalens farther from the light beam, we can clearly see the 1st order that the metalens is optimized for as the largest peak, with the 0th and -1st order as small peaks. Figure \ref{fig:waveforms}b shows a light map of the metalens, where each horizontal slice represents a waveform like in Figure \ref{fig:waveforms}a.  The black, blue, white, red and green lines mark the position of the -2,-1,0,1,2 orders, respectively, according to equation \ref{eqn:difraction_orders}.

\begin{figure}[htbp]
\centering 
\includegraphics[width=.7\textwidth]{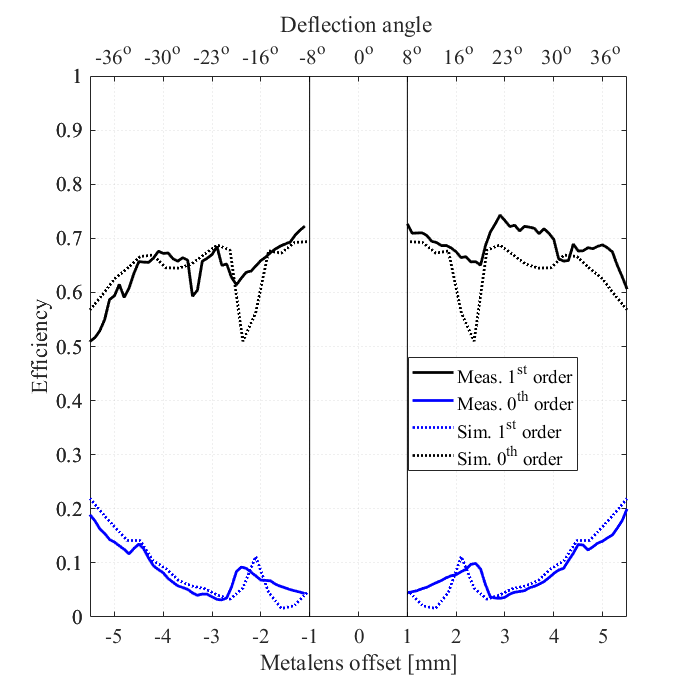}
\includegraphics[width=.7\textwidth]{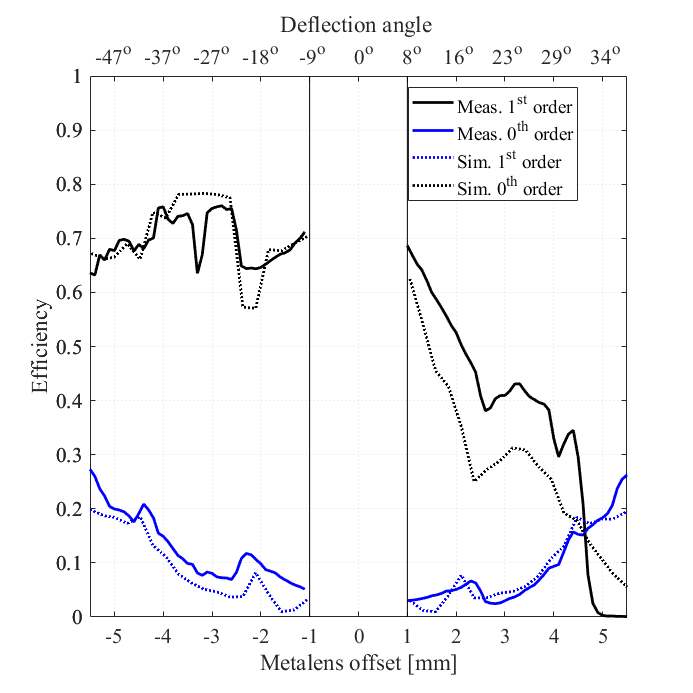}
\centering
\caption{\label{fig:data} Light transmission efficiency measured (solid lines) and simulated (dashed lines) for the 0th (blue) and 1st (black) orders, in function of the metalens offset from the light source. The top axis relates this offset to the resulting deflection angle to the first order. The region marked by horizontal lines represents the area where the orders overlap, so no distinction can be made between orders. Top: Metalens is perpendicular to the light source. Bottom: Metalens is at a 20-degree angle relative to the light source.}
\end{figure}

\begin{figure}[htbp]
    \centering
    \includegraphics[width=.7\textwidth]{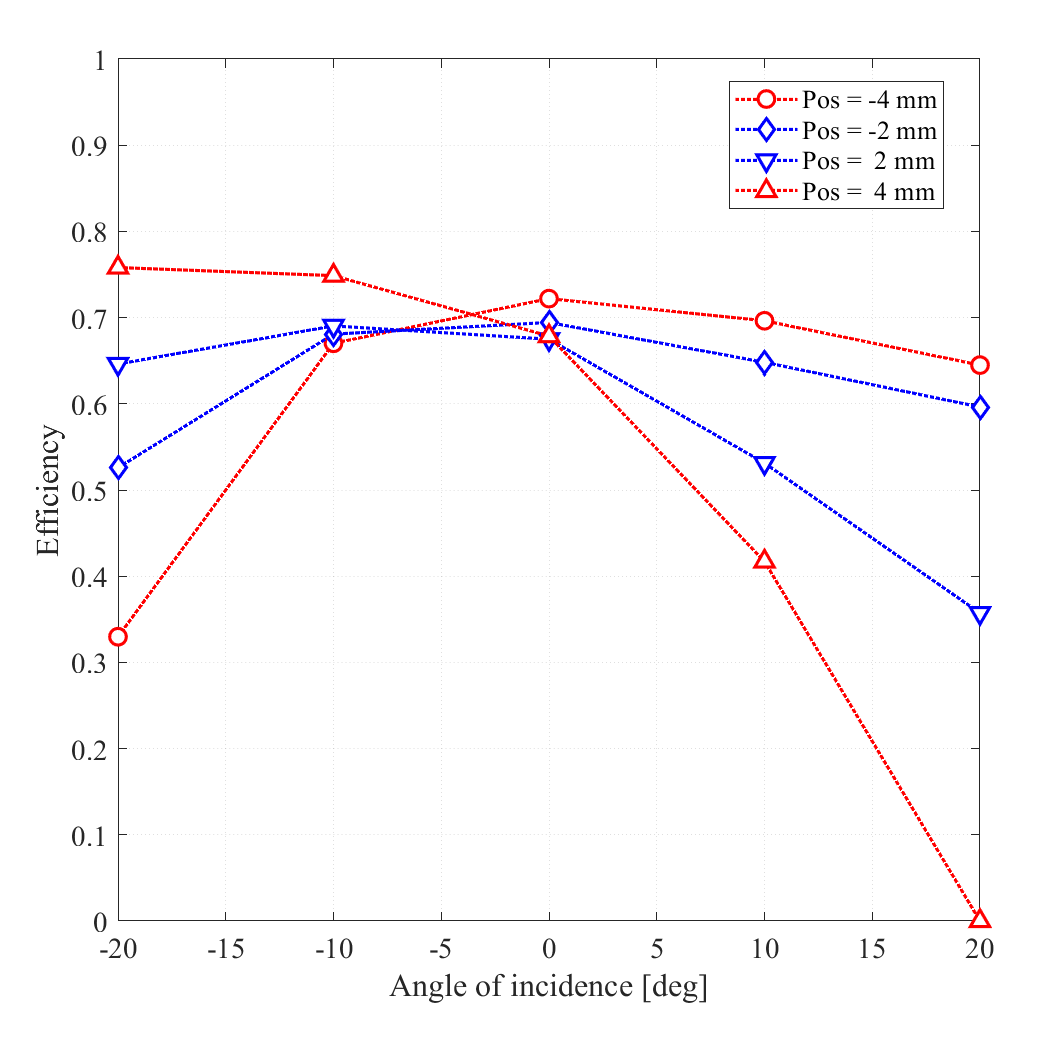}
    \caption{The light transmission efficiency of the 1st order as a function of the incidence angle. The colored lines show the four metalens offsets relative to the light source. }
    \label{fig:effvsang}
\end{figure}

The diffraction efficiency from a given order was found by taking the maximum of that order's peak, which represents the total power of that given order, and dividing by the maximum of the reference waveforms peak, which is the total incident power on the metalens. The reference waveform was taken using only the substrate used by the metalens. By finding the efficiency of a given order across the entire metalens, we can see the response of the metalens, shown in Figure \ref{fig:data}b. The measured and simulated efficiencies match quite well, giving us confidence in the simulations to produce the metalens response. We then repeated these measurements with the metalens at different incident angles from the incoming light source, shown in Figure \ref{fig:effvsang}. The efficiency of the metalens decreases with increasing angle of incidence.

\section{Discussion}

The characterization described here provides a powerful method of measuring the efficiency of metalenses across the surface of the lens and as a function of light incidence angle. Combined with the focal length and NA, these provide the full set of optical parameters needed to fully simulate the response of these devices in a particle physics detector environment. 

With the use of stepper motors, we were able to fully automate the system to characterize the light collection as a function of angle of refraction, as well as angle of incoming light. Typical measurements in the metalens community fully illuminate the lens and measure the focusing efficiency, lacking the response measurements necessary to use metalenses in particle physics detectors. Our method provides a full, low cost, and local characterization of the metalens by using the stepper motors to take automated measurements of the angular and position response. We have demonstrated the effectiveness of this characterization, as the ease of data taking allowed us to quickly optimize a metalens design for  632 nm light, used in this setup. 

With this method, we found the measured and simulated efficiency matches quite well, giving us confidence in the simulations to produce the metalens response. This is another major step in using metalenses in light collecting detectors, which would require simulations to test the use of metalenses in specific environments, light profiles, and detector geometries. Future work includes measuring the efficiency with metalenses designed for UV, VUV and visible light, where this method for characterization can be used after replacing with the appropriate metalens and light source.

\acknowledgments
Part of this research was funded by the Sloan Foundation under a 2021 Alfred P. Sloan Research Fellowship.  R. Guenette, T. Contreras are partly supported by ERC-2020-SyG-951281. Fermilab participation was supported  by a Laboratory Directed Research and Development (LDRD) grant L2021.011. J. Mart\'in-Albo is supported by the Ramón y Cajal program (ref. RYC2021-033265-I) funded by the Spanish MCIN/AEI/10.13039/501100011033 and by the EU (NextGenerationEU/PRTR).

\bibliographystyle{IEEEtran}
\bibliography{bibliography}









\end{document}